# Target tests for the ILC positron source

T. Lengler[1], D. Lott[2], G. Moortgat-Pick[1,3], S. Riemann[4]

[1] II. Institute for Theoretical Physics, University of Hamburg, Luruper Chaussee 149, 22761 Hamburg, Germany

[2] Helmholtz-Zentrum Hereon, Max-Planck-Straße 1, 21502 Geesthacht, Germany

[3] Deutsches Elektronen-Synchrotron (DESY), Notkestr. 85, 22607 Hamburg, Germany

[4] Deutsches Elektronen-Synchrotron (DESY), Plantanenallee 6, 15738 Zeuthen, Germany

# Target tests for the ILC positron source

T. Lengler[1], D. Lott[2], G. Moortgat-Pick[1,3], S. Riemann[4]

[1] II. Institute for Theoretical Physics, University of Hamburg, Luruper Chaussee 149, 22761 Hamburg, Germany

[2] Helmholtz-Zentrum Hereon, Max-Planck-Straße 1, 21502 Geesthacht, Germany

[3] Deutsches Elektronen-Synchrotron (DESY), Notkestr. 85, 22607 Hamburg, Germany

[4] Deutsches Elektronen-Synchrotron (DESY), Plantanenallee 6, 15738 Zeuthen, Germany

Abstract

The positron source is an essential component of the International Linear Collider (ILC) and is an area that poses some design challenges. One consideration is the material for the target, where the $10^{14}$ positrons per second for the ILC are generated. The potential material would need to be able to resist the high load created by the intense high energy photon beam. One of such candidates is the titanium alloy Ti-6Al-4V, for which the results of material tests with 3.5 MeV electrons are shown. The material was characterized after the irradiation by high-energy X-ray diffraction (HE-XRD) and changes caused by the irradiation to the crystal structure were studied. These tests revealed there was only minimal change in the phase fractions and crystal structure of the material under conditions as expected for the ILC.

**Introduction**

The ILC is a positron electron collider that aims to study new physics beyond the standard model. For the positron source of such an accelerator no reference exists, that can give an already well tested prototype[1]. One of the design options, the creation via high energy photons, offers the possibility for linear polarized positrons. To withstand the high thermal load generated through the photon beam and to distribute the radiation damage throughout a larger volume, the target is rotated. This results in rapid, localized heating in the target. A promising candidate material for the target is the titanium alloy Ti-6Al-4V, which offers great mechanical stability at elevated temperatures. Nevertheless, the material must be tested to study the impact of the radiation on the crystal structure of the material to determine the suitability of the material. Multiple of such tests were performed at the Mainzer Microtron (MAMI)[2]. For this particular study these tests were done using electrons with energies of 3.5 MeV, since there are no sources of high energy photons that are sufficient to test the conditions closely matched to the ILC. The material was then measured with HE-XRD to study the crystal structure and changes in the phase structure of the material.

|  | Spot | Irradiation Duration [min] | Repetition rate [Hz] | Pulse length [ms] | DT per pulse [K] | Duty factor | $T_{ave}$ at spot [C] |
|---|---|---|---|---|---|---|---|
| Sample 1 | Top | 90 | 10 | 1.5 | 335 | 0.03 | 155 |
| 200µm | Middle | 40 | 20 | 1.5 | 350 | 0.015 | 98 |
|  | Bottom | 40 | 100 | 1.0 | 240 | 0.10 | 365 |
| Sample 2 | Top | - | - | - | - | - | - |
| 250µm | Middle | 45 | 120 | 1.0 | 230 | 0.12 | 425 |
|  | Bottom | 40 | 140 | 0.5 | 170 | 0.07 | 300 |
| Sample 3 | Top | 40 | 100 | 0.5 | 170 | 0.05 | 280 |
| 500µm | Middle | 40 | 100 | 1.0 | 230 | 0.10 | 435 |
|  | Bottom | 40 | 140 | 0.5 | 160 | 0.07 | 345 |

Table 1: Shown are the irradiation parameters for all the irradiation spots.

### Experiment

To test the impact of the photon beam at the ILC on the titanium targets, they were irradiated with a 3.5 MeV electron beam at MAMI. The beam was focused to a size of 0.12 mm diameter. With this a sufficient energy deposition density could be achieved. The beam had a repetition rate of 10 to 140 Hz to study the effect of a cyclical irradiation. Pulse lengths were kept between 0.5 to 1.5 ms. With these fast repetition rates a number of cycles could be simulated, that is close to the expected amount for the target during the entire lifetime.

With the high energy material science (HEMS) beam line P07 at the synchrotron source PETRA-III at DESY operated by the Helmholtz-Zentrum Hereon there is an excellent setup for diffraction measurements, in large part due to the high-quality synchrotron radiation. The measurements were carried out at the side station of the beam-line P07, where at a photon energy of 87.1 keV the crystal structure of the samples could be studied in transmission geometry. In transmission geometry the whole depth of the sample at a given spot contributes to the diffraction pattern. Depending on the measurement series beam sizes from 300 µm x 300 µm to 100 µm x 100 µm were used. With these beam parameters an area scan of the irradiated areas was performed. For the distance between the measurement points values between 300 µm and 100 µm were used to achieve a good spatial resolution on the samples.

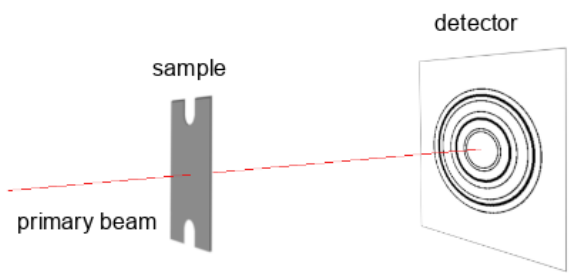

Figure 1: Schematic drawing of the setup for the experiment at the side station at the HEMS beamline at PETRA-III, DESY, operated by Hereon.

The material for the samples was the alloy Ti-6Al-4V, which is an alloy in the α+β region of the binary Ti-Al phase diagram. Therefore, fine grains of both the α and β phase will be present, which has a strong impact on the properties of the alloy.

### Samples

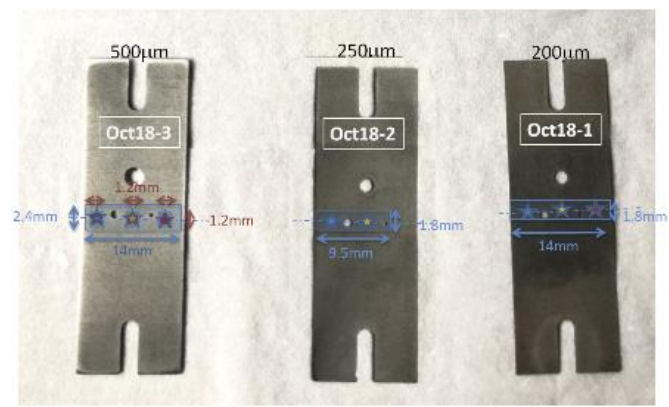

Figure 2: View of the irradiated samples with the location of the irradiation spots with the scanned areas.

A total of three Ti-6Al-4V samples as shown in Figure 2 were irradiated. They were cut from a block of Ti-6Al-4V via electro erosion with a surface of 42 mm x 14 mm and varying thickness. The thinnest of the targets had a thickness of 200 µm, which went up to 500 µm for the thickest target. There were 8 irradiations spots across these three targets. The parameters for each of the spots is listed in table 1. Additionally to the listed irradiation times, the central spot for the first sample was irradiated for ca. 3 second continuously. Marked with the blue area are the areas measured with HE-XRD, which contain all the irradiation spots.

### Results

For each of the measured areas a two-dimensional area scan was performed, where a diffraction pattern was recorded for every point. These contain valuable information about the grain size and intensity of each reflex of the crystal lattice.

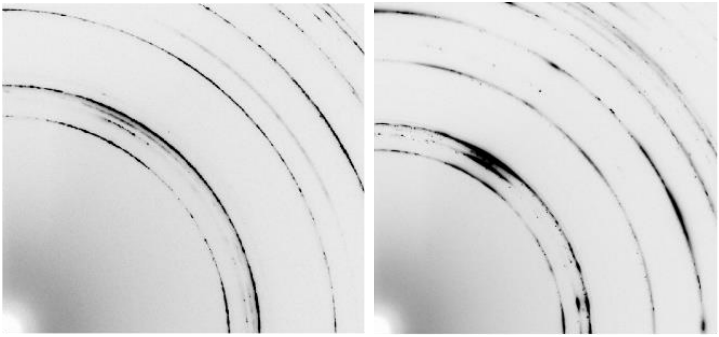

Figure 3: Top right section of a diffractogram for sample 1 outside (left) and inside (right) the irradiated area. The β phase is drastically reduced

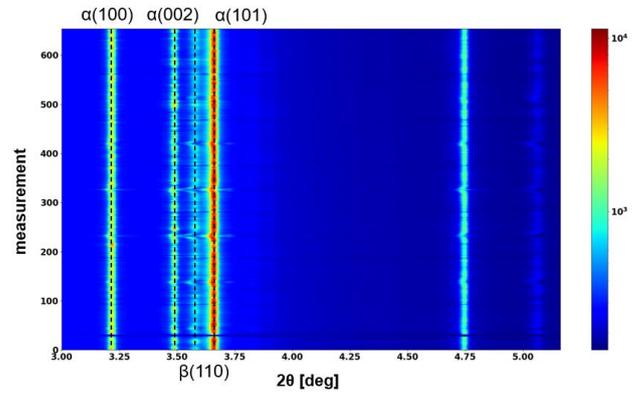

Figure 4: Visualisation of the recorded diffractograms for sample 1. Only for the continuously irradiated spot visual changes can be

For the thin plates studied in this experiment there is some texture present for the samples, which can be introduced during the manufacturing process. Therefore, the diffraction rings are mostly continuous and only some of the possible crystallite orientations are not present. No reflexes from large singular grains can be observed. For both the α and β phase there is no change from the undisturbed to the irradiated area, except for the continuously irradiated area in sample 1. Here there is nearly no more intensity present in the diffraction peaks of the β phase. However, the rings for the α phase remain continuous with no massive grain growth occurring.

**Data reduction**

For further studies these diffraction patterns were radially integrated to reduce them to a 1d diffraction pattern. While information of the orientation of the grains in the sample is lost, the averaged integrated intensity and the scattering angles can be determined and studied for each of the reflexes. In Figure 4 the change in scattering angle for both the α and β phase can be seen for the continuously irradiated spot of sample 1, as well as the decrease in intensity of the β reflexes in the very centre of the irradiation spot. This occurred for the continuously irradiated spot, but not for the spots irradiated cyclically as intended.

In the further analysis two complementary techniques were used. A quantitative phase analysis allows for the determination of the phase fractions and lattice parameters while taking the whole diffraction pattern into account. The study of singular peaks offers more detailed information of the behaviour of single peaks of a given phase.

**Quantitative phase analysis**

Using the software MAUD[2] a quantitative analysis for all the samples could be achieved, providing information about the fractions with which both phases are present in the sample.

For the spots where the sample was irradiated with the load expected at the ILC the samples did not show any signs of changes in phase fraction. Only for the continuously irradiated spot a change could be measured. Here the sample reached a temperature higher than under normal operation would be expected.

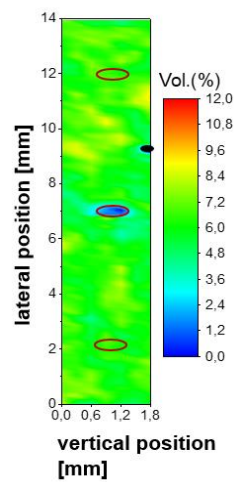

Figure 5: Results of the quantitative phase analysis for sample 1.

**Peak analysis**

The peak analysis will be presented for select peaks of the α and β phase of the titanium alloy.

For the continuously irradiated spot a change in peak parameters could be measured for the width and scattering angle of both the α and β phase, which can be seen in figure 6 for the α phase and in figure 7 for the β phase. The reduction in the scattering angle towards the beam centre means an expansion of the crystal lattice close to the beam spot.

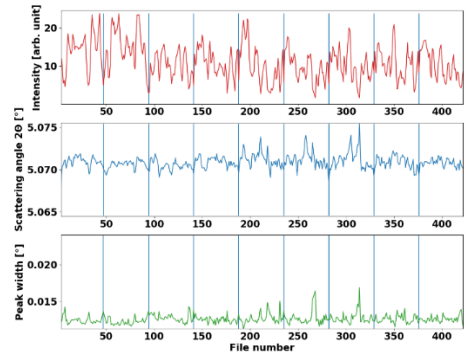

Figure 8: Results of the peak analysis of the β(200) peak for sample 3. Sectioned by the horizontal blue line are the different heights measured for

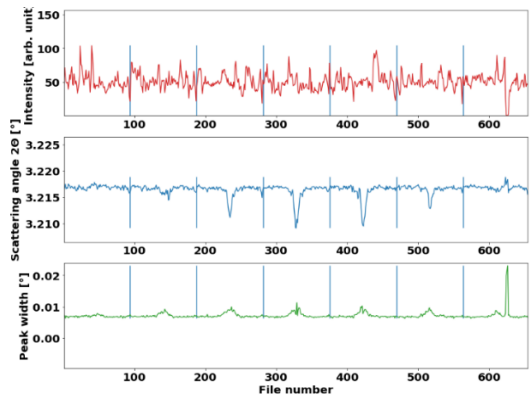

Figure 6: Results of the peak analysis of the α(100) peak for sample 1. Sectioned by the horizontal blue line are the different heights measured for

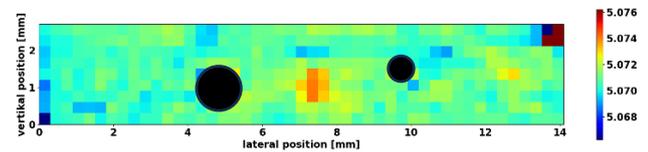

Figure 9: Visualization of the scattering angle for sample three in real space on the sample. Marked in black are the positions of the alignment holes.

**Conclusion**

Only for continuous rather than pulsed irradiation major changes could be observed. For both phases, the lattice parameters changed as well as the phase fractions for the two phases, potentially resulting in a change in properties for the alloy. Only for continuous rather than pulsed irradiation major changes could be observed. For the parameters of the normal operating conditions there is no phase transformation occurring, which could significantly alter the properties of the alloy. However, there are a minor change in the scattering angle and therefore the lattice parameter of the β phase for the most intense irradiation parameters that were tested.

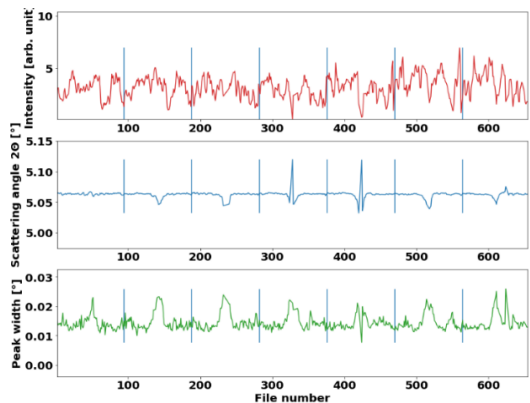

Figure 7: Results of the peak analysis of the β(200) peak for sample 1. Sectioned by the horizontal blue line are the different heights measured for the sample.

For the spots that were irradiated as expected for the ILC, there were no changes for the α phase of the alloy, which was completely stable under the irradiation. Only for the most intense irradiation parameters changes could be observed for sample 3 in the scattering angle for the β phase. These changes are much smaller than for the continuous radiation.

**Outlook**

Further experiments were conducted to study the impact of the irradiation of the material. In May 2023 additional samples could be tested at Mainz with a higher number of cycles to gauge the long-term impact of the irradiation more accurately. These can be combined with data from experiments with uniform cyclical heating to determine the impact of the localised and rapid heating as well as the displacements due to the high energy particles in contrast to the influence of the temperature increase.

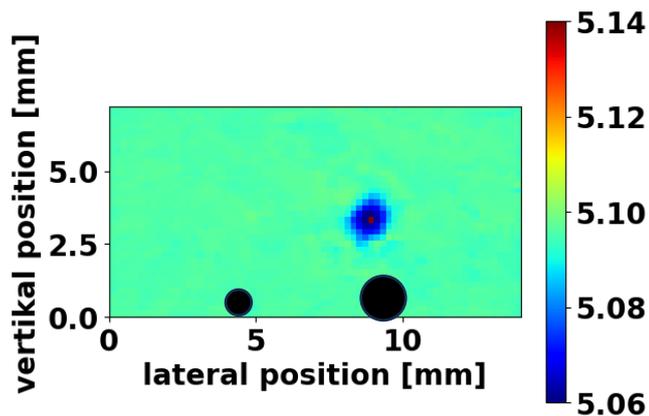

Figure 10: First results for a 500 µm thick target of the run in May. Shown is the scattering angle of the β phase. Marked in black are the positions of the alignment holes.